\documentclass[%
 aip,
 amsmath,amssymb,
 reprint,%
]{revtex4-2}
\usepackage{graphicx}
\usepackage{dcolumn}
\usepackage{bm}

\usepackage[utf8]{inputenc}
\usepackage[T1]{fontenc}
\usepackage{mathptmx}
\usepackage{etoolbox}

\makeatletter
\def\@email#1#2{%
 \endgroup
 \patchcmd{\titleblock@produce}
  {\frontmatter@RRAPformat}
  {\frontmatter@RRAPformat{\produce@RRAP{*#1\href{mailto:#2}{#2}}}\frontmatter@RRAPformat}
  {}{}
}%
\makeatother
\begin{document}
                                                                                                                                                                                                                                                                                                                                                                                                                                                                                                                                                                                                                                                                                                                                                                                                                                                                                                                                                                                                                                                                                                                                                                                                                                                                                                                                                                                                                                                                                                                                                                                                                                                                                                                                                                                                                                                                                                                                                     
\title{Rectification and bolometric   terahertz radiation detectors
based on perforated  graphene  structures exhibiting plasmonic resonant response
}
\author{V.~Ryzhii$^{1,2*}$,   C.~Tang$^{1,2}$,  
M.~Ryzhii$^{3}$, T. Otsuji$^{4,5}$ and M. S. Shur$^{6,7}$}
\address{
$^1$Research Institute of Electrical Communication,~Tohoku University,~Sendai~ 980-8577,
Japan\\
$^2$Frontier Research Institute for Interdisciplinary Sciences,
Tohoku University, Sendai 980-8578, Japan\\
$^3$School of Computer Science and Engineering, University of Aizu, Aizu-Wakamatsu 965-8580, Japan\\
$^4$ Center of Excellence ENSEMBLE3, Warsaw 01-919,\\
 Poland\\
$^5$ International Research Institute of Disaster Science, Tohoku University, Sendai 980-0845, Japan\\
$^6$Department of Electrical, Computer, and Systems Engineering,\\ Rensselaer Polytechnic Institute,~Troy,~New York~12180,\\ USA\\
$^7$ Electronics of the Future, Inc., Vienna, VA 22181-6117,\\ USA\\
*{Author to whom correspondence should be addressed: vryzhii@gmail.com}
}
\begin{abstract}
We propose and evaluate  the characteristics of the  terahertz (THz)  detectors based on perforated  graphene layers (PGLs). The PGL structures  constitute  the interdigital in-plane arrays of  the graphene microribbons (GMRs)
connected by the sets of narrow constrictions, which form the  graphene nanoribbon (GNR) bridges. 
The PGL detector operation is associated with the rectification and hot-carrier bolometric
mechanisms.  The excitation of plasmonic oscillations in the GMR-GNR arrays
can reinforce these mechanisms.
. The room temperature PGL detector  responsivity and detectivity are calculated as afunction of the radiation frequency and device structure parameters. The effects of the rectification and hot-carrier mechanisms are compared.
  The PGL THz detectors under consideration can exhibit highly competitive values of  responsivity and detectivity. 
\end{abstract}


\maketitle

\section{Introduction}
Apart from already realized graphene device structures (see the recent papers~Refs.~1-7 and the references therein), the topological structures based on  the graphene micro- and  nanoribbon (GMR and GNR) arrays, graphene nanomeshes (GNMs), and perforated
graphene layers (PGLs)  provide new opportunities to create new infrared (IR) and terahertz (THz) detectors with  elevated performance.~\cite{8,9,10,11,12,13}

\begin{figure}[b]
\centering
\includegraphics[width=8.5cm]{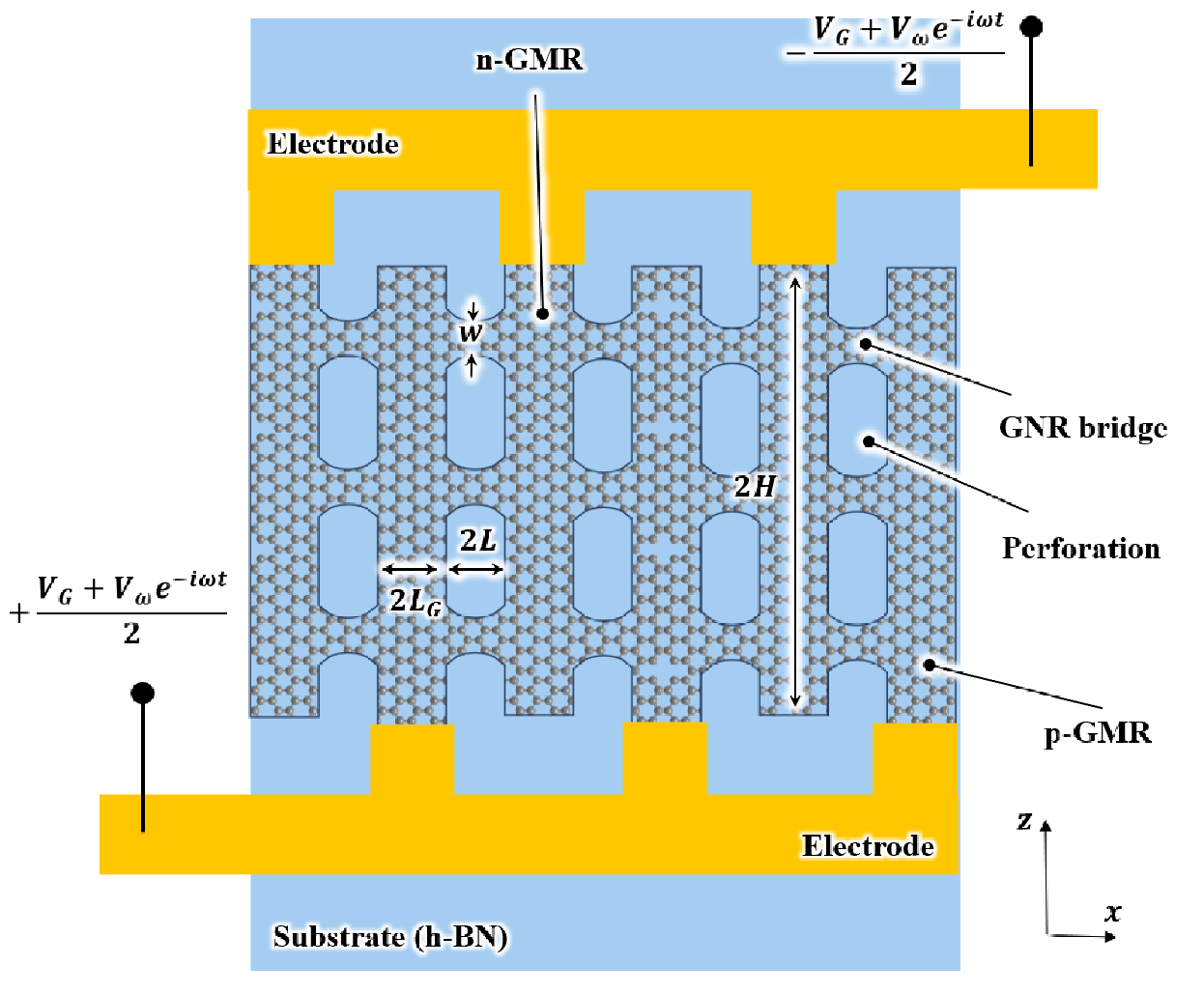}
\caption{Top view of the PGL THz detector structure.
}\label{Fig1}
\end{figure}
\begin{figure}[t]
\centering
\includegraphics[width=8.5cm]{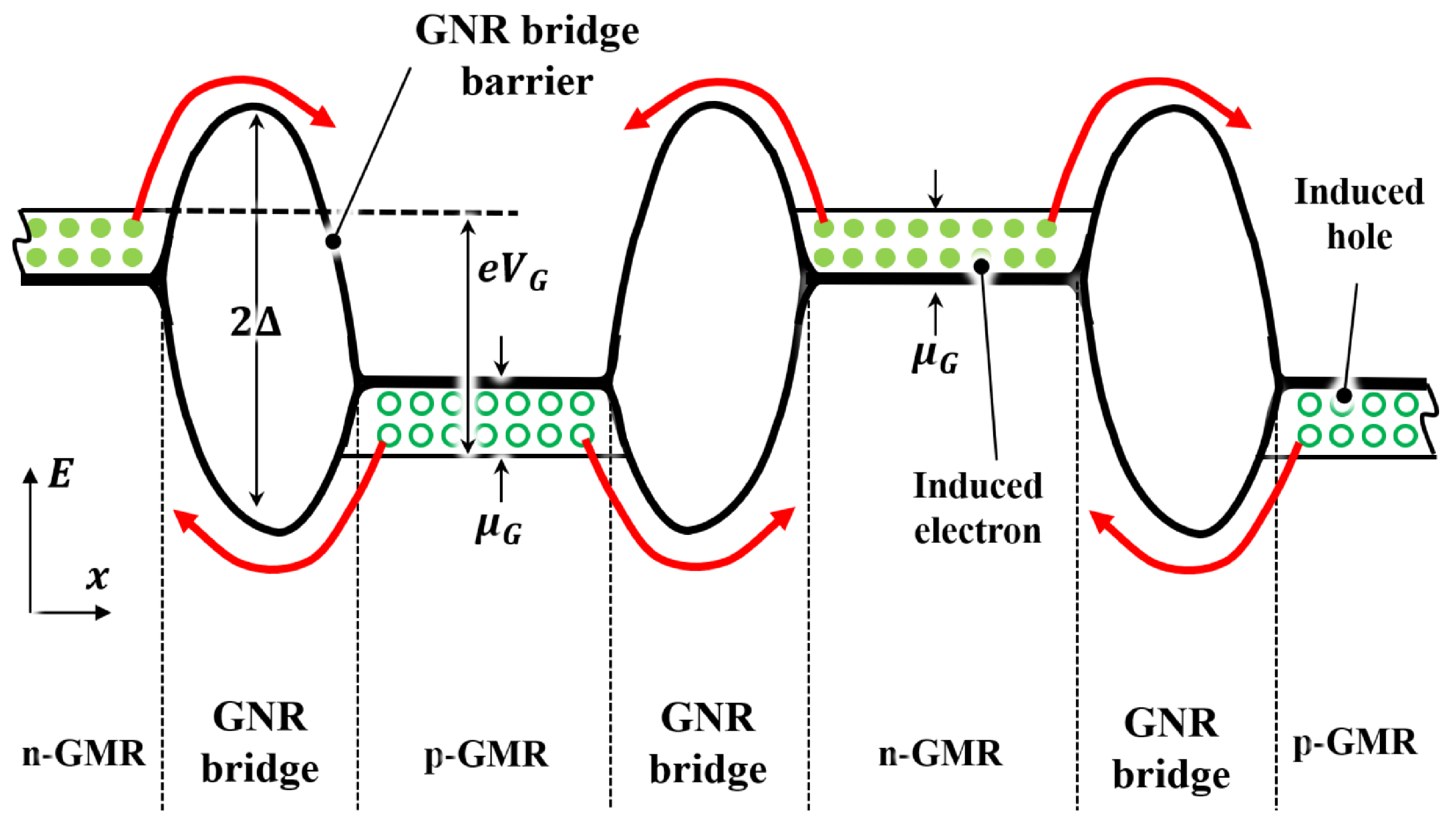}
\caption{ PGL   band diagram  at bias voltage $V_G$ .
}\label{Fig2}
\end{figure}

In this paper, we propose and  analyze the rectification and hot-carrier bolometric THz detectors
based on the periodic PGL structure. Figure~1  schematically illustrates the device's top view.
The device structure constitutes an array of the interdigital GMRs with the GNR bridges connecting the neighboring 
GMRs. These GNR bridges are the constrictions in the GL between the perforations, where the transverse quantization of the carrier energy spectra leads to the energy gap opening.~\cite{14,15,16,17} The latter results in the formation of energy
barriers for the carrier transfer between the GMRs.

Below, we calculate the room temperature  responsivity and detectivity of the PGL THz detectors accounting for
the thermionic transport through the GNR bridges and the plasmonic response
as a function of device structural characteristics and applied voltage.
We also 
estimate the speed of the detector  operation (maximum modulation frequency of the detected THz radiation).
We demonstrate that the PGL detectors can be effective in different THz systems.

\section{Device model and operation principle}

We assume that the minimal thickness of the GNRs  is chosen  to be  sufficiently small to provide a reasonable bandgap opening and, hence, a reasonable  barrier height that  effectively controls the thermionic carrier transport via the GNRs. We also assume  that the width of the  GNR bridges varies smoothly,
so that the barrier shape is close to parabolic.
One of the essential features of the PGL detectors under consideration is the dual role of the bias voltage $V_G$:
first, it induces the electrons in one pair of the GMRs and holes in another so that the n-GMRs and p-GMRs alternate and, second, stimulates the inter-GMR current and the net terminal current.

The detection of the THz radiation in such PGL detectors is enabled
by the variation in the thermionic  current of the carriers  heated by the incident THz radiation collected by an antenna (hot-carrier bolometric detection mechanism). Figure~2 shows the band diagram of a .../n-GMR/p-GNR/n-GMR/p-GMR/... PGL structure.

The opposite ends  of the interdigital GMRs (as shown in Fig.~1) are connected with the corresponding poles of the bias voltage source and the antenna pads.
The bias voltage $V_G$ induces the electron and hole charges, i.e., the two-dimensional electron and hole systems (2DESs and 2DHSs),
 in the  neighboring GMRs  
so that the n- and p-GMRs alternate. The PGL structures are assumed to be placed on  substrates, which enable sufficiently
high carrier mobilities along the GMRs, such as the 
hexagonal-BN substrates (or PGLs embedded in h-BN layers). 

The number, $2M$, of the GMRs can vary from two ($M =1$, i.e., one n-GMR and one p-GMR) to a rather large number 
($M \gg 1$). The number, $2N-1$, of the CNR bridges between each pair of  GMRs can also be different, provided that
the GNR characteristic width $W$ and the GMR length $2H$ obeys the following inequality $(2N-1)W \ll 2H $.
This implies that the perforation width is sufficiently large. The latter prevents the quantum coupling
of the electrons and holes belonging to the neighboring CNRs and provides the condition of smaller inter GMR
conductance compared to  the conductance along the GMRs.
In addition to the bias voltage $V_G$, the electron and hole densities depend on
the inter-GMR capacitance $c_G$.~

The energy barriers for the electrons and holes  in the PGLs on the h-BN substrate (or similar)
between the perforations (between the GNRs)
are large.
As a result, the electrons and holes  incident on these barriers are reflected 
 so that the inter-GMR current through the perforations is  suppressed. 
As a result, the DC and AC electron and hole currents between the neighboring GMRs
flow through the GNR bridges.

The height of the barriers in the GNR bridges  
is estimated as $\Delta = \pi\hbar\,v_W/w$. Here 
$w$ is the GNR minimal width ($w < W$ or even $w \ll W$), 
 $v_W \simeq 10^8$~cm/s is the carrier velocity in GLs, and $\hbar$ is the Planck constant.
 We assume that the  GNR width $W(x)$ {with $W(x)|_{x=0} = w$} corresponds to the
 parabolic form  of the energy barrier $\Delta(x)$, where $\Delta(x)|_{x=\pm L} =0$ and $\Delta(x)|_{x=0} = \Delta$.

The signal voltage, $V_{\omega}\exp(-i\omega t)$, produced by the incident THz radiation with the frequency $\omega$
and applied between the contacts to the GMRs, results in the spatiotemporal oscillations of the GMR potential and the electric field along the GMRs.
This results in the oscillations of the inter-GMR currents via the GNRs and the appearance of their rectified
component. 
The electric-field oscillations  also  heat  the 2DESs and 2DHSs in the respective GMRs, increasing the thermionic current through the GNRs. 
Both mechanisms of  the inter-GMR current  increase associated with the THz irradiation 
 can be substantially amplified  by the resonant excitation of the plasmonic waves along the GMRs, reinforcing  the detector's response.
These plasmonic waves 
strongly depend on the inter-GMR capacitance,~\cite{18,19}
 which is determined by  the GMR width  $2L_G$ and the inter-GMR spacing $2L$.~\cite{20,21}
Due to the in-plane configuration, the inter-GMR capacitance $c_G$ can be much smaller than the capacitance
of the gated GMRs (e.g., in the field-effect transistors). 
 
Since this capacitance is smaller than  the GMR-gate capacitance in the gated structures akin to the field-effect transistor structures,~\cite{22} the plasmonic frequencies in the PGL under consideration can fall into the THz range 
even for  relatively  long GMRs.

\section{General equations of the model}

Since the  electron and hole densities in the 2DESs and 2DHSs, respectively, under the operation conditions are sufficiently high, the electron and hole distribution functions are close to the Fermi distribution, characterized by the effective temperature and the Fermi energy. This is because the carrier-carrier interactions dominate over other carrier momentum relaxation mechanisms. The carrier effective temperature is, generally, different from lattice temperature $T_0$, in particular, due the carrier heating by the impinging radiation.   

Considering that the GNRs form  near-parabolic energy barriers,
and using
 Landauer-Buttiker formula~\cite{23} (see also, e.g., Refs.~24 and 25) applied to the one-dimensional transport through the GNRs,  one can arrive at  the following approximate expression for
the net current $J$:

\begin{eqnarray}\label{eq1}
J \simeq M(2N-1)J^{GNR},
\end{eqnarray}

\begin{eqnarray}\label{eq2}
J^{GNR} = \frac{8eT}{\pi\,\hbar}\exp\biggl(\frac{-\Delta+\mu_G}{T}\biggr)\sinh\biggl(\frac{\eta\,eV_G}{2T}\biggr).
\end{eqnarray}
Here $T$ and $T_0$ are  the carrier  effective temperature and the lattice temperature (in the energy units), respectively, 
 $\mu_G = e\sqrt{{\overline V}_GV_G}$ is the  Fermi energy in the respective GMRs induced by the bias voltage, ${\overline V}_G = (\pi\,c_G\hbar^2v_W^2/2e^3L_G)$ is the characteristic voltage, 
  $c_G= [(\kappa_S+1)/4\pi^2]{\overline c}_G$ is the inter-GMR capacitance, $\kappa_S$ is the substrate dielectric constant,  ${\overline c}_G$ is a slow function of  the $L_G/L$ ratio,~\cite{21,22} and
  $\eta \lesssim 1$ is a coefficient describing details of  barrier shape modification (in the following we set for brevity $\eta = 1$). 
Due to the heating of the holes by   injecting electrons  and  the heating of the electrons  by injecting  holes,
the DC carrier temperature ${\overline T}$  exceeds the ambient temperature $T_0$.

\section{Dark current}

As follows from Eqs.~(1) and (2),
the terminal current in the absence of irradiation (dark current) ${\overline J}$ is given by   
 
\begin{eqnarray}\label{eq3}
{\overline J} \simeq 8M(2N-1)\frac{e^2{\overline T}}{\pi\,\hbar}
\exp\biggl(\frac{-\Delta +\mu_G}{{\overline T}}\biggr)\sinh\biggl(\frac{\eta\,eV_G}{2{\overline T}}\biggr).
\end{eqnarray}

 The DC  component of the carrier effective temperature ${\overline T}$ 
 is found accounting for the energy balance between the power, ${\overline P} = {\overline J} V_G$ received by the 2DESs and 2DHSs in the GMRs due to  the DC  currents and the power, with $4MHL_G\Sigma_G{\overline R}$ being  transferred to the lattice (see Appendix A),

\begin{eqnarray}\label{eq4}
{\overline P} = 4MHL_G\Sigma_G{\overline R}.
\end{eqnarray}
 Here $\Sigma_G = c_GV_G/e$ is the carrier density  electrically induced in the GMRs with
 $c_G$ being the inter-GMR capacitance (per unit of their length).  
 The ${\overline T} -V_G$ and ${\overline J}- V_G$ relations, obtained solving Eqs.~(3) and (4) with  Eq.~(A1), 
 are shown in Fig.~3. We assumed $T_0 = 25$~meV and the following PGL structural parameters: $2N-1 = 5$, $\tau_{\varepsilon} = 20$~ps, $\Delta = 200 - 300$~meV 
 [$w \simeq (6 - 10) $~nm], $2H = 1.0~\mu$m, $2L_G = 60$~nm, and $2L = 40$~nm.  

\begin{figure}[t]
\centering
\includegraphics[width=8.5cm]{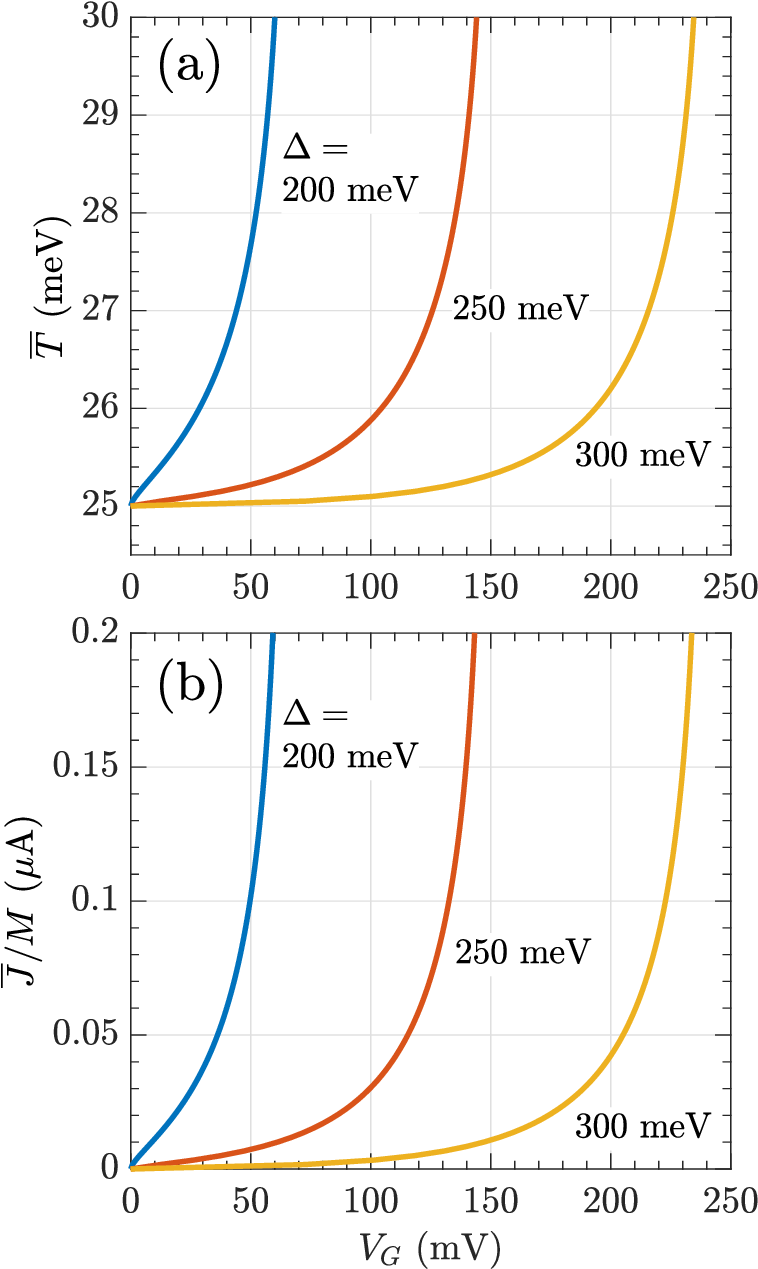}
\caption{ Voltage dependences of  (a) DC carrier temperature ${\overline T}$ and (b) dark current 
${\overline J}$ (normalized by number of GMR pairs $M$) for PGL structures with different GNR barrier heights $\Delta$.
}
\label{Fig3}
\end{figure}
As seen from Fig.~(3), ${\overline T}$ and ${\overline J}$ steeply rise with increasing $V_G$.
The   behavior of the ${\overline T} - V_G$ and ${\overline J} - V_G$ characteristics is determined by 
 parameter $\Theta_N$ proportional to
the number of the GNR bridges per unit of the GMR length $(2N-1)/2H$. 
When $\Theta_N$ is sufficiently large, $d{\overline T}/dV_G$ and  $d{\overline J}/dV_G$
turn to infinity  at certain critical voltages (hot-carrier thermal breakdown).~\cite{26} 
In agreement with Fig.~3, in the PGLs with the above parameters, the critical thermal breakdown voltage is estimated as ${\tilde V}_G\simeq (60 - 230)$~mV.

In the PGL structures with a moderate parameter $\Theta_N$, the ${\overline T} - V_G$ and ${\overline J} - V_G$ characteristics are monotonically rising.~\cite{26}

\section{THz photocurrent and plasmonic resonant response}

The AC component $\Delta T_{\omega}$ is found
using  the pertinent equation governing the carrier heating  associated with the THz radiation 

The signal voltage, $V_{\omega}$, produced by the impinging THz radiation  generates the potential spatiotemporal oscillations of the GMR potential $\varphi_{\omega}^+ = \varphi_{\omega}^+(t,z)$ (in the p-type GMRs)
and $\varphi_{\omega}^- = \varphi_{\omega}^-(t,z)$ (in the n-type GMRs), producing the rectified component of the DC current between the GMRs and the bolometric component associated with  the variation in the carrier effective temperature  $\Delta T_{\omega}
= T- {\overline T}$. We derive the carrier temperature variation $\Delta T_{\omega}$ 
using the linearized signal version of the carrier energy balance in the form

\begin{eqnarray}\label{eq5}
P_{\omega} = 4MHL_G\Sigma_G R_{\omega}
\end{eqnarray}
with $R_{\omega} = \Delta T_{\omega}/\tau_{\varepsilon}$.

Accounting for both the rectified and bolometric contributions,
the variation in the terminal current, caused by the THz irradiation (THz photocurrent),
can be presented as

\begin{eqnarray}\label{eq6}
\Delta J_{\omega } \simeq 
\frac{1}{2}\frac{\partial^2{\overline  J}}{\partial V_G^2}\biggr|_{{\overline T}}
\langle |\varphi_{\omega}^+-\varphi_{\omega}^-|^2\rangle
 +
\frac{\partial{\overline  J}}{\partial {\overline T}}\biggr|_{V_G} 
\langle|\Delta T_{\omega}|\rangle
\end{eqnarray}
 with

\begin{eqnarray}\label{eq7}
\Delta T_{\omega} = \frac{\tau_{\varepsilon}}{\Sigma_G}
\biggl[
\sigma_{\omega}^{GNR}\frac{\langle |\varphi_{\omega}^+-\varphi_{\omega}^-|^2\rangle}{4HL_G}
+
\sigma_{\omega}^{GMR}\langle\biggl|\frac{\partial\varphi_{\omega}^{\pm}}{\partial z}\biggr|^2 \rangle\biggr].
\end{eqnarray}

Here $\sigma_{\omega}^{GNR}$ and  $\sigma_{\omega}^{GMR}$ 
are the real parts of the differential conductance of the GNRs (neglecting the transit time effect) and   the GMR   longitudinal AC Drude conductivity (described by Eqs.~(A5) and (A6) in Appendix A), respectively, and 
 the symbol
 $\langle...\rangle$
implies the averaging over the GMR length and the THz signal period.
The first and second terms in the right-hand side of Eq.~(6) describe
the rectified and bolometric components of the THz photocurrent, respectively.
Equation~(7) accounts for the fact that $\Delta T_{\omega}$ is determined by  both   the injection of hot carriers from one GMR to another (proportional to $|\varphi_{\omega}^+ -\varphi_{\omega}^-|^2$)
  and with the Joule heat due to the GMR longitudinal conductivity (proportional to $|\partial \varphi_{\omega}^{\pm}/\partial z|^2$). 

Accounting for the spatial distributions of the AC potential $\varphi_{\omega}^+=\varphi_{\omega}^+(z)$
(along the p-GMRs) and   $\varphi_{\omega}^- =\varphi_{\omega}^-(z)$ (along the n-GMRs),  presented
by Eqs.~(B2) - (B4) in Appendix B, 
for the  quantities $\langle|\varphi_{\omega}^+ - \varphi_{\omega}^-|^2\rangle$ and
$\langle|\partial \varphi^{\pm}/\partial z|^2\rangle$ in Eq.~(7) we obtain

\begin{eqnarray}\label{eq8}
\langle|\varphi_{\omega}^+ - \varphi_{\omega}^-|^2\rangle
=\frac{1}{4}\frac{|1 +\sin \gamma_{\omega} \cos\gamma_{\omega}/\gamma_{\omega}|}{|\cos\gamma_{\omega} -\gamma_{\omega} \sin\gamma_{\omega}|^2} V_{\omega}^2
\end{eqnarray}
and

\begin{eqnarray}\label{eq9}
\langle\biggl|\frac{\partial \varphi^{\pm}}{\partial z}\biggr|^2\rangle
=\frac{1}{4}\frac{\gamma_{\omega}^2}{H^2}\frac{|1 +\sin^2\gamma_{\omega}-\sin \gamma_{\omega} \cos\gamma_{\omega}/\gamma_{\omega}|}{|\cos\gamma_{\omega} -\gamma_{\omega} \sin\gamma_{\omega}|^2} V_{\omega}^2
\end{eqnarray}
Here
$\displaystyle \gamma_{\omega} 
  = 
 \frac{ \pi\sqrt{\omega(\omega+i\nu)}}{2\Omega}$ and $\displaystyle \Omega = \frac{e^{3/2}}{H\,\hbar}\sqrt{\frac{\pi\sqrt{{\overline V}_GV_G}\,L_G}{4c_G}}$
are  the normalized plasmonic wave number  and the
  characteristic plasmonic frequency, respectively.

Using Eqs.~(7) - (9), we arrive at the following formula relating the carrier temperature variation
 $\Delta T_{\omega}$ and the amplitude of the signal voltage, $V_{\omega}$, produced by the impinging  
 THz radiation between the GMR ends:

\begin{eqnarray}\label{eq10}
\Delta T_{\omega} \simeq \frac{e\tau_{\varepsilon}\nu}{4V_G}\biggl(\delta^{GNR}|1 +\zeta_{\omega}|
%
+\frac{\omega\,|1- \zeta_{\omega}|}{\sqrt{\nu^2+\omega^2}}
\biggr)\,
\Pi_{\omega}V_{\omega}^2.
\end{eqnarray}
 Here
 
 \begin{eqnarray}\label{eq11}
 \Pi_{\omega} = \frac{1}{|\cos\gamma_{\omega} -\gamma_{\omega} \sin\gamma_{\omega}|^2}
 \end{eqnarray}
is the factor associated with the plasmonic resonances in the electrically coupled  n-type  and p-type GMRs,
$\delta^{GNR} = \sigma^{GNR}/c_GH\nu$,
and $\zeta_{\omega}=\sin \gamma_{\omega} (\cos\gamma_{\omega}/\gamma_{\omega} - \sin\gamma_{\omega})$.

Parameter $\delta^{GNR}$
is proportional to the  exponential factor [see Eq.~(A5)], which is small in practical range of the bias
voltages. This corresponds to the smallness of the GNR conductance compared with the GMR conductance.
Hence, the first term in the brackets in Eq.~(10) can be disregarded.

Using Eqs.~(6), (7), (8), and (10), we obtain the following expression for the rectified, 
$\Delta J_{\omega }^R$  and bolometric $\Delta J_{\omega}^B$  components of the net  THz photocurrent $\Delta J_{\omega}^R + \Delta J_{\omega}^B $:

\begin{eqnarray}\label{eq12}
\Delta J_{\omega }^R \simeq 
\frac{e^2{\overline J}}{32{\overline T}^2}
|1 +z_{\omega}|\Pi_{\omega} V_{\omega}^2
\end{eqnarray}
 and
 
\begin{eqnarray}\label{eq13}
\Delta J_{\omega}^B
\simeq  
\frac{{\overline  J}(1 +F)}
{{\overline T}}
\frac{e\tau_{\varepsilon}\nu}{4V_G}
\frac{\omega\,|1- \zeta_{\omega}|}{\sqrt{\nu^2+\omega^2}}
\Pi_{\omega}V_{\omega}^2,
\end{eqnarray}
respectively, where 
 $F = \exp[(\Delta - \mu_G -eV_G/2)/{\overline T}]$.

\section{Contribution of rectified  and bolometric  mechanisms to PGL detector responsivity}

Keeping in mind that the upper bound of $V_{\omega}^2$ and the power, $P_{\omega}$, received by the detector antenna
are related as $V_{\omega}^2 =16\pi^2P_{\omega}/c$, 
where $c$ is the speed of light in vacuum, for the detector responsivity $R_{\omega}^R =
\Delta J_{\omega}^R/P_{\omega}$, we arrive at

\begin{eqnarray}\label{eq14}
R_{\omega}^R \simeq 
%
{\overline R}|1 +z_{\omega}|\Pi_{\omega}
\end{eqnarray}
and  
\begin{eqnarray}\label{eq15}
R_{\omega}^B = {\overline R}\, B\,\frac{\omega\,|1- \zeta_{\omega}|}{\sqrt{\nu^2+\omega^2}}\Pi_{\omega},
\end{eqnarray} 
where 

\begin{eqnarray}\label{eq16}
{\overline R} = \frac{\pi^2}{2c}\frac{{e^2\overline J}}{{\overline T}^2}
\end{eqnarray}
is the PGL detector characteristic responsivity and
the factor

\begin{eqnarray}\label{eq17}
B = 8\tau_{\varepsilon} \nu(1+F)({\overline T}/eV_G)
\end{eqnarray}
dependent on the bias voltage $V_G$ can be called  the bolometric factor.

Figure~4 shows the ${\overline R} - V_G$ relations calculated using Eq.~(16) invoking the ${\overline T} - V_G$ and ${\overline J}-V_G$ characteristics found above and corresponding to Fig.~3.

\begin{figure}[t]
\centering
\includegraphics[width=8.5cm]{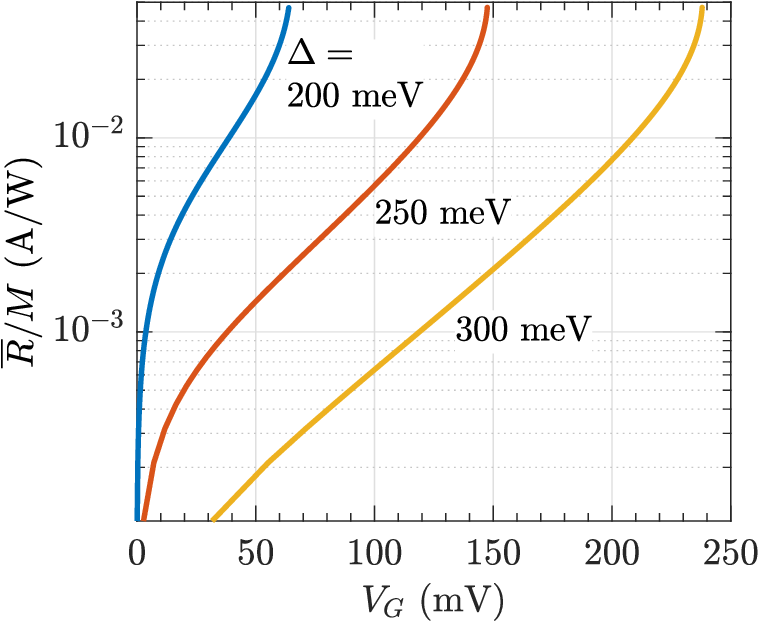}
\caption{Characteristic  responsivity ${\overline R}$ (normalized by number of GMR pairs $M$) vs bias voltage $V_G$
for PGL structures with different GNR barrier heights $\Delta$ and  the same parameters as in Fig.~3}
\label{Fig4}
\end{figure}

The characteristic responsivity ${\overline R}$ steeply rises with increasing bias voltage $V_G$, particularly at moderate values of the GNR barrier height $\Delta$. However, the maximum responsivity value
at chosen $\Delta$ is limited by the values of   $V_G$ being less than the critical voltage, ${\tilde V}_G$ at which
$d{\overline T}/d V_G$ and $d{\overline J}/d V_G$ might turn to infinity
(see Sec.~IV), beyond which
the steady-state current flow can be unstable.~\cite{26} Therefore, the voltage range $V_G \sim {\tilde V}_G$
is  unsuitable for the PGL detector operation due to the excessive noise (see below).

As an example, the   estimated  characteristic responsivity ${\overline R}$, which is common for both rectification and bolometric detection mechanisms, for the main parameters used above and the particular case $M = 5 - 10$, $\Delta = 250$~meV, and  $V_G = 120$~mV, is about  ${\overline R} \simeq (0.05 - 0.10)=$~A/W [${\overline T} \simeq 27$~meV and ${\overline J} \simeq (0.5 - 1.0)~\mu$A]. Equation~(15) shows 
that
the  bolometric responsivity $R_{\omega}^B$ can substantially exceed ${\overline R}$ by factor $B \gg 1$.

Both $R^R_{\omega}$ and   $R^B_{\omega}$ are proportional to the frequency-dependent plasmonic factor $\Pi_{\omega}$. This factor can exhibit pronounced peaks at the resonant frequencies 
$\omega_0 \simeq 1.72\Omega/\pi$ and $\omega_{n>0} \simeq 2\Omega(1+ 1/\pi^2n)$ with $n = 1, 2,3,...$~\cite{18}
provided the quality factor of the plasmonic oscillations $Q \sim \Omega/\nu \gg 1$.   
 Considering that for the above parameters, the plasmonic  frequency $\Omega/2\pi \simeq 0.69$~THz.
 At shorter GMR lengths, the plasmonic frequency $\Omega/2\pi$ can be markedly higher ($\Omega \propto H^{-1}$).
  Using  Eqs.~(13) - (16), for the ratio $R^B_{\omega}/ R^R_{\omega}$ we obtain

\begin{eqnarray}\label{eq18}
\frac{R_{\omega}^B}{R_{\omega}^R} = B\,\frac{\omega}{\sqrt{\nu^2+\omega^2}} \biggl|\frac{1- \zeta_{\omega}}{1+ \zeta_{\omega}}\biggr|.
\end{eqnarray} 
Since $B \gg 1$, the right-hand side of Eq.~(18) is large except, possibly, the 
range of low frequencies $\omega\ll \nu, \Omega$.  This implies that the bolometric detection mechanism prevails in the most interesting frequency range, particularly in the THz range.

Figure~5 shows the frequency dependences of the normalized responsivities 
 $R^R_{\omega}/{\overline R}$ and   $R^B_{\omega}/{\overline R}$ calculated
 using Eqs.~(14) - (16) for the above parameters and $\nu = (0.5 - 1.5)$~ps$^{-1}$.
As seen, a decrease in the carrier collision frequency $\nu$ (an increase in the carrier mobility in the GMR) results in a marked rise of the detector responsivity and the sharpening of the plasmonic resonances.

Since the plasmonic frequency $\Omega$ increases with the bias voltage ($\Omega \propto V_G^{1/4}$, in line with Ref.~20), the plasmonic resonances and, hence, the PGL detector spectral characteristics are voltage-controlled.

However, the efficiency of the bolometric mechanism drastically drops in the case of detection of the modulated THz signals with the modulation frequency $\omega_M/2\pi > 1/2\pi\,\tau_{\varepsilon} $.~\cite{27}
 In contrast, 
the rectification mechanism can still be efficient at the modulation frequencies far beyond $1/2\pi\,\tau_{\varepsilon}$, i.e., up to sub-THz frequencies.

\begin{figure}[t]
\centering
\includegraphics[width=8.5cm]{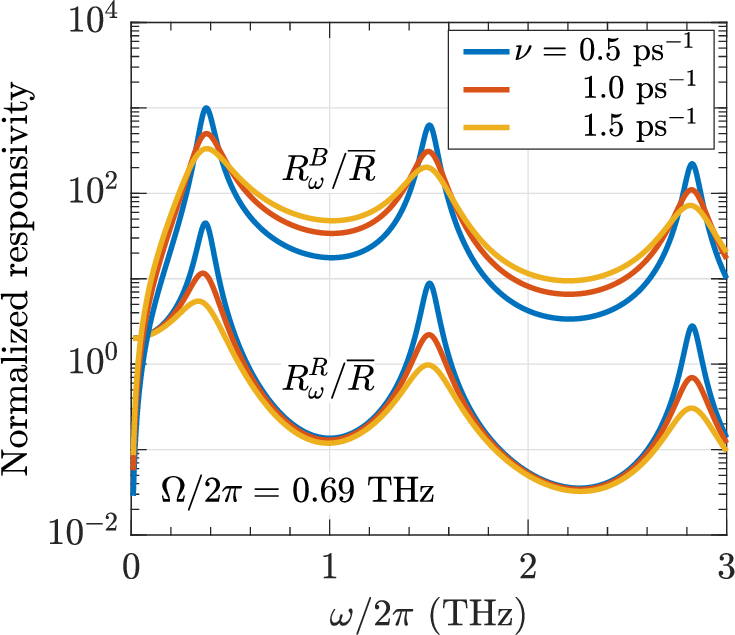}
\caption{Normalized responsivities $R_{\omega}^R/{\overline R}$ and $R_{\omega}^B/{\overline R}$ vs signal frequency $\omega/2\pi$ for PGL structure with $\Delta = 250$~meV and different values of carrier collision frequency $\nu $ at $V_G = 120$~mV
(other parameters are  the same parameters as in Fig.~3.}
\label{Fig5}
\end{figure}

 \section{PGL detector detectivity}

We estimate  the PGL detector characteristic detectivity, $D_{\omega}$,  in the units $\sqrt{\rm Hz}/W$  (the inverse noise equivalent power) 
associated with the dark-current noise  and   Nyquist-Johnson noise accounting for that with the pertinent noise currents are  $i_{DC}^2 \propto 4e{\overline J}$ and
$i^2_{NJ} \propto 4e{\overline T}(d{\overline J}/dV_G$
using the following definition:

\begin{eqnarray}\label{eq19}
 D_{\omega} = \frac{R_{\omega}^R+ R_{\omega}^B}
{\sqrt{4e{\overline J}}+\sqrt{4{\overline T} d{\overline J}/dV_G}}\nonumber\\
 = {\overline D}\biggl(|1+z_{\omega}|
+ |1-z_{\omega}|\frac{\omega}{\sqrt{\nu^2+\omega^2}}\biggr)\Pi_{\omega}.  
\end{eqnarray}
Here

\begin{eqnarray}\label{eq20}
{\overline D} = \frac{{\overline R}}
{\sqrt{4e{\overline J}}}\frac{1}{\biggl(1+ \sqrt{\displaystyle\frac{{\overline T}}{e{\overline J}}  \frac{d{\overline J}}{dV_G}}\biggr)}\nonumber\\
 = \frac{\pi^2e^{3/2}}{4c}\frac{\sqrt{{\overline J}}}{{\overline T}^2}
\frac{1}{\biggl(1+ \sqrt{\displaystyle\frac{{\overline T}}{e{\overline J}}  \frac{d{\overline J}}{dV_G}}\biggr)} 
\end{eqnarray}
is the characteristic detectivity.

\begin{figure}[t]
\centering
\includegraphics[width=8.5cm]{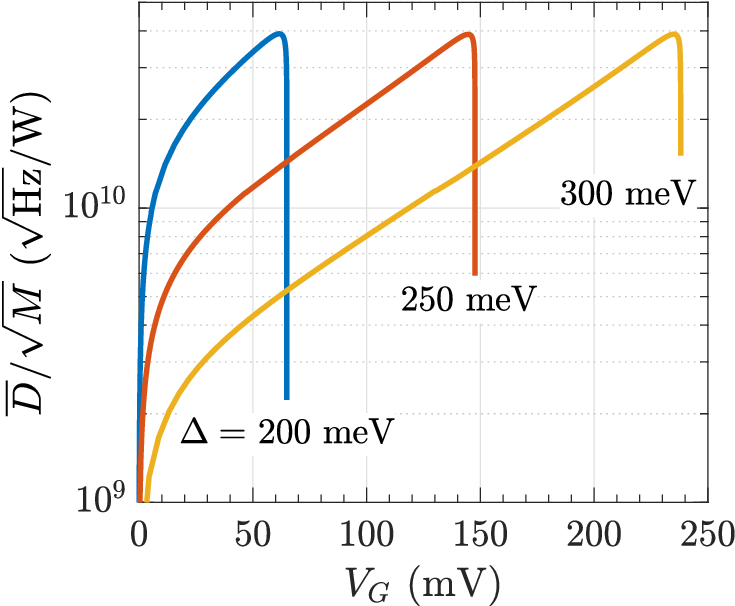}
\caption{Characteristic detectivity ${\overline D}$ (normaized by factor $\sqrt{M}$) versus  bias voltage $V_G$ for the same PGL structure parameters as in Fig.~4.}
\label{Fig6}
\end{figure}

In the range of relatively low bias voltages, where ${\overline T} \sim T_0$, Eq.~(20) yields 

\begin{eqnarray}\label{eq21}
{\overline D} \simeq \frac{{\overline R}}
{\sqrt{4e{\overline J}}}\frac{\sqrt{2}}{1+\sqrt{2}}
\simeq 1.48 2\frac{e^{3/2}}{c}\frac{\sqrt{{\overline J}}}{ T_0^2}.
\end{eqnarray}
Substituting the data obtained in Sec.~VI, namely,  $M = 5 - 10$, $V_G = 120$~mV,  ${\overline J } \simeq (0.5 - 1.0)~\mu$A ,and ${\overline R} \simeq
(0.05 - 0.10)$~A/W, we find ${\overline D} \simeq (5.2 - 7.4)\times 10^{10}\sqrt{{\rm Hz}}$/W. 

Due to $D_{\omega} \propto \Pi_{\omega}$, the detectivity frequency dependence is similar to that
of the responsivity shown in Fig.~5.
Considering that the bolometric factor $B$ and the plasmonic factor $\Pi_{\omega}$, can be fairly large,
the detector detectivity $D_{\omega}$ can substantially exceed the latter value (up to two orders of magnitude).

However, in the PGL detectors with relatively large $\Theta_N \propto ((2N-1)/2H$ (like that characterized by the parameters used in Figs.~3 and 4) at $V_G$ approaching  the critical voltage ${\tilde V}_G$, $d{\overline J}/dV_G$  can tend to infinity so that ${\overline D}$ drastically rolls-off tending to zero. As a result, the ${\overline D} - V_G$ characteristics can exhibit  maxima. 
The existence of this maximum  is attributed to the   competition between the dark-current noise  and the  Nyquist-Johnson noise. Figure~6 shows the voltage dependences of the characteristic detectivity
${\overline D}/\sqrt{M}$ calculated for the PGLs with different GNR barrier heights $\Delta$.
These plots clearly demonstrate a nonmonotonic behavior of the ${\overline D} - V_G$ relations.
A drastic drop of the detectivity when $V_G$ approaches to a certain value (the threshold voltage 
${\tilde V}_G$) is due to a jump of the   Nyquist-Johnson noise, which, in turn, is associated with the related rise of the differential conductance $d{\overline J}/d V_G$.

The  ${\overline D} - V_G$ relation in the PGL structures characterized by a moderate parameter $\Theta_N$ still exhibits a maximum with, however, fairly smooth roll-off at large $V_G$.

According to Eqs.~(14) and (15), the responsivities $R_{\omega}^R$ and
$R_{\omega}^B$ can substantially exceed ${\overline R}$ at the plasmonic resonances.
Hence, the  detectivities associated with the rectified  mechanism
and, particularly with the bolometric mechanism  can be much larger than the  value obtained in the latter estimate for ${\overline D}$ and ${\overline D} B$.  
Since ${\overline J}$ is proportional to the number of GMR pairs $M$,
the PGL detector  responsivity and detectivity increase with increasing $M$, being proportional to $M$ and $\sqrt{M}$, respectively.

\section{Comments}

Above, we limited our consideration by relatively moderate bias voltages (see, Figs.~3 - 5).
The point is that in the PGL structures with a sufficiently large number of GNR bridges $(2N-1)$,
the voltage dependence of ${\overline J}$ can  sharply increase with $\sigma_{\omega}^{GNR} \propto d{\overline J}/d V_G$ turning to infinity.~\cite{26} 
This implies that the ${\overline T} - V_G$ and ${\overline J}- V_G$ 
characteristics can have the 
S-shape 
at the bias voltages exceeding certain critical values, depending on $(2N-1)$ and some other structural parameters.
In this situation,  the low-carrier-temperature and low-current regime under consideration above (corresponding to the sub-critical bias voltages)   becomes unstable.   The transition from low-temperature to high-temperature regime, i.e., the switching between the low- and upper
branches of  the ${\overline T} - V_G$ and ${\overline J} - V_G$ characteristics
can be accompanied by excessive  Nyquist-Johnson noises decreasing the PGL detector performance.

Generally, the electron or hole transit via the  energy barrier is associated
with the thermo-assisted tunneling. The contribution of the electrons and holes with the energies below the parabolic barrier top is determined by the "tunneling" temperature,~\cite{28} which for the parabolic barrier of width $2L$ is estimated as
$\Theta_{tunn} \simeq (\hbar\,v_W/2\pi\,L)$. 
%

For the PGL structures   under consideration with $2L = 40$~nm, the latter formula yields the following estimate:
$\Theta_{tunn} \simeq 5$~meV (i.e., $\Theta_{tunn}$ is markedly smaller than $T_0$ and ${\overline T}$ ). 
This justifies disregarding  the  thermo-assisted tunneling compared with the thermionic processes used in  the model used above.

The thermionic model of the inter-GMR injection based on Eqs.~(1) - (3) and their consequences is valid
if $\mu_G + eV_G/2 < \Delta$. This leads to the inequality $V_G < {\tilde V}_G =(
\sqrt{4\Delta/e + {\overline V}_G} - \sqrt{{\overline V}_G})^2$. For the parameters 
assumed in the calculations, ${\tilde V}_G \simeq (300 - 475)$~mV, i.e., substantially exceeds the bias voltages $V_G$ corresponding to the above results. 

As follows from the above data, the uncooled  PGLs exploiting the hot-carrier bolometric mechanism can exhibit elevated responsivity $R_{\omega} \propto {\overline R}$ and detectivity $D_{\omega} \propto {\overline D}$ (i.e., low noise equivalent power NEP $=D_{\omega}^{-1} \propto {\overline D}^{-1}$), especially at plasmonic resonances and sufficiently large number of the GMRs $M$). This enables the PGLs
to be  competitive  surpassing the existed room-temperature THz microbolometer THz detectors (see, for example, Refs.~1 - 3).
In particular, the plots in Figs.~4 and 5 for $M = 10$ correspond to max~$R_{\omega} \simeq 10$~A/W.
The values of the noise equivalent power for the PGLs evaluated above can be NEP $\ll 100$~pW  much lower than those of the best uncooled bolometric THz detectors.~\cite{1,2,3}

The proposed PGL detectors are  based on GMR/GNR arrays with properties compatible  with current fabrication capabilities. High-resolution fabrication of sub-10-nm GNRs with atomically smooth edges has already been demonstrated experimentally using squashed carbon nanotubes,~\cite{30} as well as  other lithographic and bottom-up approaches including 
 techniques such as CVD
combined with electron-beam lithography or self-assembly
followed by transfer.~\cite{31,32,33,34} 

\section{Conclusions}

We proposed and analyzed the detection mechanisms (rectification and hot-carrier bolometric) in the PGL THz detectors  and evaluate their responsivity and detectivity. As shown, the responsivity and detectivity of these detectors is primarily determined by the bolometric mechanism except for the detection of the THz radiation modulated in the sub-THz range. Increasing the number of  GMR pairs and the
carrier mobility leads to rising detector responsivity and detectivity.
The PGL  detectors can exhibit fairly high values of room-temperature responsivity and detectivity in the THz range, especially, at the plasmonic resonant frequencies.

\section*{Acknowledgments}
The work was supported by
JST ALCA-NEXT (Grant No.24021835),
NEDO (Grant No.20020912),
Murata Foundation (Grant No.AN24322), and
Iketani Foundation (Grant No.0361181-A), JSPS (KAKENHI Grant No. 21H04546), Japan,
and by AFOSR (Contract No. FA9550-19-1-0355),
USA.

\section*{Author declaration}
The authors have no conflicts to declare.

\section*{Data availability}
The data that support the findings of this study are available
within the article.

\section*{Appendix A. Energy balance in the GMRs}
\setcounter{equation}{0}
\renewcommand{\theequation} {A\arabic{equation}}

The energy balance  determining  the relationship between the DC effective temperature ${\overline T}$
and the bias voltage $V_G$ is described by Eq.~(4).
Considering that at room temperatures and not too high bias voltages, the main carrier energy relaxation mechanism is associated primarily with optical phonons (see, e.g.,, Refs.~35-40), 
for the values of power, ${\overline P}$
and $P_{\omega}$, transferred to the lattice [used in Eqs.~(4) and (5)], we obtain

\begin{eqnarray}\label{eqA1}
{\overline R} \simeq \frac{\hbar\omega_0}{\tau_{\varepsilon}}
\biggl(\frac{T_0}{\hbar\omega_0}\biggr)^2
\biggl[\exp\biggl(\frac{\hbar\omega_0}{T_0} - \frac{\hbar\omega_0}{{\overline T}}\biggr)-1\biggr],
\end{eqnarray}
 where $\hbar\omega_0$ and $\tau_{\varepsilon} = \tau_0\exp(\hbar\omega_0/T_0)(T_0/\hbar\omega_0)^2$ are  the optical phonon energy in GMRs and the energy relaxation time of "warm" carriers, respectively, and $\tau_0$ is the time of optical phonon spontaneous emission (which is in the sub-picosecond range).
 Other energy relaxation mechanisms, such as the disorder-assisted electron scattering and plasmon-mediated 
processes,~\cite{40,42,43} as well as the carrier heating/cooling at the GMR side contacts,~\cite{27,43,44,45,46} which in the PGLs under consideration can be crucial at higher bias voltages~\cite{26} are disregarded.
In Eq.~(A1)  we also neglected the contribution of the interband recombination-generation processes associated with optical phonons 
because the Fermi levels are well above the Dirac point in the n-type GMRs and well below this point in the p-type GMRs (compared with Refs.~37, 47, and 48). 
The heating of the 2DESs and 2DHSs in the GMRs is associated with the energy brought by the electrons
injected into the p-GMRs and the holes injected into the n-GMRs. Since such an energy (per one carrier)
is equal to $eV_G$,
we have 
 
\begin{eqnarray}\label{eqA2}
 {\overline P} = eV_G {\overline J}.
\end{eqnarray}

 The variation of the carrier effective temperature $\Delta T_{\omega}$ obeys Eq.~(7) with

\begin{eqnarray}\label{eqA3}
 R_{\omega} \simeq \frac{\Delta T_{\omega}}{\tau_{\varepsilon}}
\end{eqnarray}
 and
 
\begin{eqnarray}\label{eqA4}
 P_{\omega} \simeq 
\sigma_{\omega}^{GNR}\langle |\varphi_{\omega}^+-\varphi_{\omega}^-|^2\rangle
+
4HL_G\sigma_{\omega}^{GMR}\langle\biggl|\frac{\partial\varphi_{\omega}^{\pm}}{\partial z}\biggr|^2 \rangle.
\end{eqnarray}
Here

\begin{eqnarray}\label{eqA5}
\sigma_{\omega}^{GNR} = \frac{4(2N-1)e^2}{\pi\hbar}\exp\biggl(\frac{-\Delta+\mu_G}{{\overline T}}\biggr)
\cosh\biggl(\frac{eV_G}{2{\overline T}}\biggr)
\end{eqnarray}

and
\begin{eqnarray}\label{eqA6}
\sigma_{\omega}^{GMR} = \frac{e^2\mu_G}{\pi\hbar^2}\frac{\nu}{(\nu^2+\omega^2)}\
\end{eqnarray}
are the real parts of the GNR and GMR conductances with  $\nu$ being the frequency of the carrier collisions in the GMRs on acoustic phonons and impurities.
The symbol
 $\langle...\rangle$
implies the averaging over the GMR length and the THz signal period.


\section*{Appendix B. Plasmonic oscillation stimulated by THz irradiation}
\setcounter{equation}{0}
\renewcommand{\theequation} {B\arabic{equation}}

The signal voltage results in the spatiotemporal variations in the GMR potentials $\varphi_{\omega}^{\pm}$,
which oscillate with the frequency $\omega$ and vary along the GMRs (in the $z$-direction) and, hence, in the longitudinal AC electric fields $\partial\varphi_{\omega}^{\pm}/\partial z $.
 In the sufficiently perfect GMRs,
these variations can constitute  plasmonic oscillations (the standing waves with the wave vectors directed along the GMRs, i.e., in the $z$-direction~\cite{18}). We describe the plasmonic oscillations  by the standard hydrodynamic equations~\cite{49}  governing the carrier transport
along the GMRs coupled with  the Poisson equations for self-consistent potential.
Solving the linearized versions of these equations with the boundary conditions

\begin{eqnarray}\label{eqB1}
 \varphi_{\omega}^{\pm}(z)|_{z=\pm H} = \pm V_{\omega}/2,\qquad
[\partial  \varphi_{\omega}^{\pm}(z)/\partial z]|_{z=\mp H} =0,
\end{eqnarray} 
dictated by the PGL structure geometry and circuitry, we obtain  
the following expression for 
the potential drop, $\varphi_{\omega}^{+}(z) - \varphi_{\omega}^{-}(z)$, across the GNRs  (compared with the corresponding formulas in Refs.~18, 59, and 60:.

\begin{eqnarray}\label{eqB2}
\varphi_{\omega}^{\pm}(z) =  \pm\frac{[\cos(\gamma_{\omega}z/H)
\mp (z/H)
\gamma_{\omega}\sin\gamma_{\omega}]}{(\cos\gamma_{\omega} -\gamma_{\omega} \sin\gamma_{\omega})} \frac{V_{\omega}}{2},
\end{eqnarray}
so that

\begin{eqnarray}\label{eqB3}
\varphi_{\omega}^{+}(z) - \varphi_{\omega}^{-}(z) =  \frac{\cos(\gamma_{\omega}z/H)
} {(\cos\gamma_{\omega} -\gamma_{\omega} \sin\gamma_{\omega})}V_{\omega}
\end{eqnarray}
and

\begin{eqnarray}\label{eqB4}
\frac{\partial \varphi_{\omega}^{\pm}(z)}{\partial z}  =  \mp\frac{\gamma_{\omega}}{H}\frac{[\sin(\gamma_{\omega}z/H)\pm \sin\gamma_{\omega}]
} {(\cos\gamma_{\omega} -\gamma_{\omega} \sin\gamma_{\omega})}V_{\omega}.
\end{eqnarray}

Here
$\displaystyle \gamma_{\omega} 
  = 
 \frac{ \pi\sqrt{\omega(\omega+i\nu)}}{2\Omega}$ and $\displaystyle \Omega = \frac{e}{H\,\hbar}\sqrt{\frac{  \pi\mu\,L_G}{4c_G}}$
are  the plasmonic wave number  and the
  characteristic frequency of the plasmonic oscillations.

\end{document}